\documentclass[12pt, a4paper]{article}
\usepackage{a4wide,amssymb,amsmath,amscd,graphicx}
\usepackage[all]{xy}
\newcommand{\sw}{\stackrel{\star}{\wedge}}

\begin{document}

\topmargin -2pt
\headheight 0pt

\topskip 0mm \addtolength{\baselineskip}{0.20\baselineskip}

\vspace{5mm}

\begin{center}

{\Large \bf Coordinate Dependence of Chern-Simons Theory\\
on Noncommutative $AdS_3$} \\

\vspace{10mm}

{\sc Ee Chang-Young}${}^{ \dag, }$\footnote{cylee@sejong.ac.kr},
{\sc Daeho Lee}${}^{  \dag,  }$\footnote{dhlee@sju.ac.kr},
and {\sc Youngone Lee}${}^{  \ddag,  }$\footnote{youngone@daejin.ac.kr}
\\

\vspace{1mm}

 ${}^{\dag}${\it Department of Physics, Sejong University, Seoul 143-747, Korea}\\

${}^{\ddag}${\it Department of Physics, Daejin University, Pocheon, Gyeonggi 487-711, Korea
}\\

\vspace{10mm}
{\bf ABSTRACT}
\end{center}

\noindent
We investigate the coordinate dependence of noncommutative theory by
studying the solutions of noncommutative $U(1,1)\times U(1,1)$
Chern-Simons theory on $AdS_3$
in the polar and rectangular coordinates.
We assume that only the space coordinates are noncommuting.
The two coordinate systems are equivalent only up to first order
in the noncommutativity parameter $\theta$.
We investigate the effect of this non-exact equivalence
between the two coordinate systems in two cases,
a conical solution and a BTZ black hole solution,
using the Seiberg-Witten map.
In each case,
the noncommutative solutions in the two coordinate systems
obtained from the corresponding same commutative solution
turn out to be different even in the first order in $\theta$.

\vfill

\noindent

\thispagestyle{empty}

\newpage

\section{Introduction}
\label{secIntro}
\setcounter{footnote}{0}


Physics in nonconcommutative spacetime has long been studied
\cite{Doplicher:1994tu,Lukierski:1991pn}
since Snyder introduced the notion of quantized spacetime \cite{Snyder:1946qz}.
Among many proposed models, the most common commutation relation (called canonical)
between coordinates is
\begin{equation}
\label{moyal_nc}
[\hat{x}^\alpha,\hat{x}^\beta]=i \theta^{\alpha\beta},
\end{equation}
where $\theta^{\alpha\beta}=-\theta^{\beta\alpha}$ are constants.
After this canonical noncommutativity was introduced
in the string theory context \cite{Connes:1997cr,Seiberg:1999vs},
it became the mainly studied commutation relation
for physics in noncommutative spacetime.

This commutation relation resembles
 the fundamental commutation relation of quantum physics.
Inspired by Weyl quantization in quantum mechanics \cite{Weyl:1927},
a theory on the canonical noncommutative spacetime\footnote{
In this paper, we only deal with space-space noncommutativity, and time is a commuting
coordinate throughout the paper.
Thus we use the terms (noncommutative) space and (noncommutative) spacetime interchangeably.}
can be reinterpreted
to another theory on the commutative spacetime
in which a product of any two functions on the original noncommutative
spacetime is replaced with a deformed ($\star$) product of the functions on
the commutative spacetime, the Moyal product \cite{Groenewold:1946kp}:
\begin{equation}
\label{moyalprd}
(f\star g)(x)\equiv \left.\exp\left[\frac{i}{2}\theta^{\alpha\beta}\frac{\partial}
{\partial x^{\alpha} }\frac{\partial}{\partial y^{\beta} }\right] f(x)g(y)\right|_{x=y} .
\end{equation}
Most of the analyses for noncommutative physics
are performed
by using the Moyal product on the commutative space
instead of being treated on noncommutative spaces directly.

What if we use a different coordinate system instead of the canonical coordinate system given by
\eqref{moyal_nc}? We expect that the commutation relations for the two coordinate systems
 would not be exactly equivalent
to each other. Would then the physics described in these two coordinates systems be the same?
We are used to take general covariance for granted.
 General covariance in ``a noncommutative space"\footnote{Here, we put the quotation mark
 since it is not clear at the moment whether we have to treat
 coordinate systems with different commutation relations of ``a given space"
 as different noncommutative spaces.
 }
   would mean the
equivalence among different coordinate systems.
However, as we mentioned above different coordinate systems in ``a noncommutative space"
generally have different
commutation relations which are not exactly equivalent.
Therefore we expect that coordinate transformations among different coordinate systems
would not yield the same physics contradicting the usual notion of general covariance.
Seiberg \cite{ns2005} has already pointed out that general covariance would be broken
in  theories with emergent spacetime  among which model theories on noncommutative spaces are
also included.
In this paper, we focus on this issue:
general covariance on a noncommutative space vs. non-exact equivalence between noncommutative coordinate
systems.
In order to check this, we compare the solutions of
$U(1,1)\times U(1,1)$ noncommutative Chern-Simons theory
in the rectangular and polar coordinates in 3-dimensional AdS noncommutative spacetime.


Gauge theory on the canonical noncommutative spacetime
has been well established using the Seiberg-Witten map \cite{Seiberg:1999vs}.
The Seiberg-Witten map is the consistency requirement for
a noncommutative gauge transformation of a gauge
theory living on a noncommutative spacetime
 to be equivalent to a gauge transformation of an ordinary gauge theory
living on a commutative spacetime.
Using this equivalence of the Seiberg-Witten map, one can find the
corresponding noncommutative gauge fields in terms of
given ordinary gauge fields. The corresponding noncommutative
gauge transformation can be found likewise.

For the three dimensional gravity, it has been well known
that it is equivalent to a Yang-Mills theory with
the Chern-Simons(CS) action in three dimensional spacetime
\cite{Achucarro:1987vz,Witten:1988hc}.
Thus using the Seiberg-Witten map
the noncommutative extension of 3D gravity-CS equivalence was studied
in \cite{Grandi:2000av,Banados:2001xw,Cacciatori:2002gq}.
Based on these works, Pinzul and Stern \cite{Pinzul:2005ta}
obtained noncommutative  $AdS_3$ vacuum and conical solution
using the Seiberg-Witten map.
Rather recently, this method was applied to the rotating BTZ black hole case\footnote{
Before this, the non-rotating BTZ black hole case had been investigated in \cite{bdrs04}
in a different set-up of geometrical framework.}
 in \cite{Kim:2007nx}
with commutation relation of $[\hat{r},\hat{\phi}]= i \theta$.

 For the four dimensional gravity, there is no such known equivalence relation
between gravity and gauge theory in four dimensional spacetime.
However, using the Poincar\'{e} gauge theory approach of Chamseddine \cite{Chamseddine:2000si}
a noncommutative Schwarzschild black hole solution was first obtained in \cite{ctz08}
using the Seiberg-Witten map.
Likewise, the charged black hole solutions in 4D were obtained in \cite{ms08,Chaichian:2007dr}.

In our previous work \cite{EDY:20081}, we studied the rotating BTZ black
hole in a noncommutative polar coordinates with the commutation relation\footnote{
This is equivalent to
$
\label{r2phi}
[\hat{r}^2,\hat{\phi}]=2i\theta .
$
}
\begin{equation}
\label{ncr2}
[\hat{r},\hat{\phi}]= i \theta  \hat{r}^{-1},
\end{equation}
which is different from the one used in \cite{Kim:2007nx} and is
equivalent to the canonical relation \eqref{moyal_nc} up to first order
in $\theta$.
In this paper, we study the rotating BTZ black hole case with the
canonical commutation relation $[x,y]= i \theta$,
 and compare it with our previous result \cite{EDY:20081}.
Then we again obtain the conical solution on $AdS_3$ in the noncommutative polar coordinates
with the commutation relation \eqref{ncr2}  and
compare it with the one obtained in \cite{Pinzul:2005ta}.
The results exhibit their dependence on a chosen coordinate system.


The paper is organised as follows.
In section \ref{secDiff},
we consider some aspects related with the Seiberg-Witten map
and then investigate the difference between the commutation relations
in the polar and rectangular coordinates.
In section \ref{secBTZ},
we obtain the noncommutative BTZ solution with the canonical
commutation relation of noncommutative rectangular coordinates, then
 compare it with the result in the noncommutative polar coordinates
 obtained in \cite{EDY:20081}.
 In section 4, we get the conical solution of noncommutative
 $AdS_3$ in the noncommutative polar coordinates, and compare it with the
 previously obtained solution by Pinzul and Stern \cite{Pinzul:2005ta}
 in which the canonical commutation relation of the rectangular coordinates
 was used.
We conclude with discussion in section 5.

\section{Different noncommutativity and Seiberg-Witten map}
\label{secDiff}

Here, we begin with reviewing
the Seiberg-Witten map and study related aspects
by treating the same map in ``a noncommutative spacetime"
with different commutation relations.
After that we show
how these noncommutativities are different
in the two following perspectives,
coordinates as operators and the
Moyal product as a deformed product from twist.

\subsection{ Seiberg-Witten map in different coordinates}

The Sieberg-Witten map matches ordinary gauge fields $\mathcal{A}$
on a commutative spacetime with noncommutative gauge fields  $\hat{\mathcal{A}}$
on a noncommutative spacetime such that an ordinary gauge transformation of $\mathcal{A}$
is equivalent to a noncommutative gauge transformation of $\hat{\mathcal{A}}$ \cite{Seiberg:1999vs}:
\begin{eqnarray}
\label{SWe}
\hat{\mathcal{A}}(g\cdot \mathcal{A}\cdot g^{-1}-\partial g\cdot g^{-1})
=\hat g*\hat{\mathcal{A}}*\hat g^{-1}-\partial \hat g* \hat g^{-1},
\end{eqnarray}
where  $*$ denotes the Moyal product,
$g,\hat g$ are elements of gauge groups for the ordinary and noncommutative gauge theories, respectively.
The above equation can be solved to first order in $\theta$ as follows.
\begin{eqnarray}
\label{Aswef}
&&\hat{\mathcal{A}}_{\gamma}(\mathcal{A})
\equiv \mathcal{A}_{\gamma}+\mathcal{A}'_{\gamma}
=\mathcal{A}_{\gamma}-\frac{i}{4}\theta^{\alpha\beta}
\{ \mathcal{A}_{\alpha},\partial_{\beta}\mathcal{A}_{\gamma}+\mathcal{F}_{\beta\gamma}
\}, \\
\label{lswef}
&&\hat{\lambda}(\lambda,\mathcal{A})
\equiv\lambda+\lambda'
= \lambda+\frac{i}{4}\theta^{\alpha\beta}
\{ \partial_{\alpha}\lambda,\mathcal{A}_{\beta}
\},
\end{eqnarray}
where $\hat{\lambda}$ and $\lambda$ are  noncommutative and
ordinary infinitesimal gauge transformation parameters.
We note that there are two important factors
in the derivation of the solution \eqref{Aswef} and \eqref{lswef}.
One is knowing of the explicit form of the Moyal product up to first order in $\theta$,
and the other is the coordinate independence of noncommutativity  parameter $\theta$
being used in the Moyal product.
One would no longer get the same form of  solution for Eq. \eqref{SWe}
in the cases of coordinate dependant noncommutativity  parameters.

Generally one obtains different solutions of the Seiberg-Witten equation
for different coordinate systems.
To see this let us consider a coordinate transformation $\varphi$
between two coordinate systems $\{x^\alpha\}$ and $\{z^a\}$,
say,
$\varphi: x^\alpha\rightarrow z^a\equiv z^a(x^\mu)$.
Then a Seiberg-Witten solution $\hat{\mathcal{A}}_c(z)$
in the coordinate system $\{z^a\}$
can be rewritten in terms of $\hat{\mathcal{A}}_\alpha(x)$,
the corresponding solution of the Seiberg-Witten equation
in the coordinate system $\{x^\alpha\}$:
\begin{eqnarray}
\label{diffSW}
\hat{\mathcal{A}}_c(z) &=&
 \mathcal{A}_{c}(z)
 -\frac{i}{4}\tilde\theta^{ab}
\{ \mathcal{A}_{a}(z),\partial_{b}\mathcal{A}_{c}+\mathcal{F}_{bc}
\}\nonumber\\
&=&
\frac{\partial x^\gamma}{\partial z^c}\mathcal{A}_\gamma(x)
-\frac{i}{4}\tilde\theta^{ab}
\left\{ \frac{\partial x^\alpha}{\partial z^a}\mathcal{A}_{\alpha}(x),
\frac{\partial}{\partial z^b}
\left(\frac{\partial x^\gamma}{\partial z^c}\mathcal{A}_{\gamma}\right)
+\frac{\partial x^\beta}{\partial z^b} \frac{\partial x^\gamma}{\partial z^c}\mathcal{F}_{\beta\gamma}
\right\}\nonumber\\
&=&\frac{\partial x^\gamma}{\partial z^c}
\left(\hat{\mathcal{A}}_\gamma(x)
+\frac{i}{4}\theta^{\alpha\beta}
\{ \mathcal{A}_{\alpha},\partial_{\beta}\mathcal{A}_{\gamma}+\mathcal{F}_{\beta\gamma}
\}\right)\nonumber\\
&&
-\frac{i}{4}\tilde\theta^{ab} \frac{\partial x^\alpha}{\partial z^a}
\left\{\mathcal{A}_{\alpha}(x),
\frac{\partial^2 x^\beta}{\partial z^b\partial z^c}\mathcal{A}_{\beta}+
\frac{\partial x^\beta}{\partial z^b} \frac{\partial x^\gamma}{\partial z^c}
\left(\partial_{\beta}\mathcal{A}_{\gamma}+\mathcal{F}_{\beta\gamma}\right)
\right\},\nonumber
\end{eqnarray}
where $\tilde\theta^{ab}$ denote noncommutativity parameters
assumed in the coordinate system $\{z^a\}$.
This can be reexpressed to show the difference between the two Seiberg-Witten solutions
in $\{z^a\}$ and  $\{x^\alpha\}$ coordinate systems,
\begin{eqnarray}
\label{diffSW2}
\hat{\mathcal{A}}_c(z)-\frac{\partial x^\gamma}{\partial z^c}\hat{\mathcal{A}}_\gamma(x)
&=&
-\frac{i}{4}\tilde\theta^{ab}
\left(\frac{\partial x^\alpha}{\partial z^a}\right)
\left(\frac{\partial^2 x^\beta}{\partial z^b\partial z^c}\right)
\{\mathcal{A}_\alpha,\mathcal{A}_\beta\}
\nonumber\\
&&
+\frac{i}{4}
\left(\frac{\partial x^\gamma}{\partial z^c}\right)
\left(
\theta^{\alpha\beta}
-\frac{\partial x^\alpha}{\partial z^a} \frac{\partial x^\beta}{\partial z^b}\tilde\theta^{ab}
\right)
\{ \mathcal{A}_{\alpha}(z),\partial_{\beta}\mathcal{A}_{\gamma}+\mathcal{F}_{\beta\gamma}\},
\end{eqnarray}
up to first order in $\theta$.
The first term on the right-hand side vanishes when the transformation $\varphi$ is linear, i.e.,
$\frac{\partial^2 x^\beta}{\partial z^b\partial z^c}= 0$.
When $\frac{\partial^2 x^\beta}{\partial z^b\partial z^c}\neq 0$,
the solution $\hat{\mathcal{A}}_\mu(\mathcal{A})|_z$
in the coordinate system $\{z^a\}$
is different from  $\hat{\mathcal{A}}_\mu(\mathcal{A})|_x$
obtained in the coordinate system $\{x^\mu\}$.
The second  term vanishes when the two noncommutativity parameters,
$\theta^{\alpha\beta}$ and $\tilde{\theta}^{ab}$, are related
as if they are tensors\footnote{
The noncommutativity parameter $\tilde\theta^{ab}$
in the polar coordinate system we use in this paper
and the canonical one $\theta^{\alpha\beta}$
satisfy this relation up to first order in $\theta$.
In fact, if the two Moyal products in the two coordinate systems are equal up to first order in $\theta$,
then one can show that this condition holds always regardless of the ordering problem.
},
$\theta^{\alpha\beta}
=\frac{\partial x^\alpha}{\partial z^a} \frac{\partial x^\beta}{\partial z^b}\tilde\theta^{ab}$.
Although the vanishing condition for the second term does not hold in general,
our polar noncommutativity parameter  $\theta /r$ in \eqref{ncr2}
and the canonical noncommutativity parameter $\theta$ in \eqref{nccan}
satisfy this condition. However, the transformation from the rectangular to the polar coordinates is not linear.
Thus, the first term does not vanish and as we shall see this difference will
yield different results for the rectangular and polar coordinate systems.

\subsection{Coordinates as operators}

In the following two subsections,
we compare the aspects of noncommutativity in the polar and  rectangular
coordinate systems especially in using the Seiberg-Witten map.
Since we consider only space-space noncommutativity
in three dimensional spacetime in this paper,
it is sufficient to compare the two sets of coordinate operators
 $(\hat{x},\hat{y})$ and $(\hat{r},\hat{\phi})$.

In the rectangular coordinate system,
the commutation relation is given in the canonical form:
\begin{equation}
\label{nccan}
[\hat{x},\hat{y}]= i \theta .
\end{equation}
When the two sets of coordinate operators
are related by the corresponding classical relation
which is not linear,
for example ($x\rightarrow r \cos\phi, ~y\rightarrow r\sin\phi$),
we face the ordering ambiguity if we want to express one set of coordinates
in terms of other set of coordinates.
Moreover, for the maps between functions of the operators,
like a solution $\hat{\mathcal{A}}(x, y)$ of the Seiberg-Witten equation on commutative space
which corresponds to $\mathcal{A}(\hat x, \hat y)$ on noncommutative space,
the ambiguity becomes severe.

In \cite{EDY:20081} it was shown that
the commutation relation between polar coordinates
is equivalent to the above canonical commutation relation
up to first order in $\theta$.
The commutation relation chosen there was the relation \eqref{ncr2} which is equivalent to
\begin{equation}
\label{ncrdouble}
[\hat{r}^2,\hat{\phi}]=2i\theta .
\end{equation}
To see how the above commutation relation and
the canonical one \eqref{nccan} is related,
we assume that the usual map $(x,y)\rightarrow (r,\phi)$
between the rectangular and polar coordinates holds
in this noncommutative space,
\begin{eqnarray}
\label{rec-pol}
\hat{x}=\hat{r}\cos\hat{\phi}, ~~\hat{y}=\hat{r}\sin\hat{\phi}.
\end{eqnarray}
Using the commutation relation  $[\hat{\phi},\hat{r}^{-1}]=i\theta \hat{r}^{-3}$
deduced from \eqref{ncrdouble} one gets:
\begin{eqnarray}
\label{r2xy}
\hat{x}^2+\hat{y}^2 := \hat{r} ( \hat{r}-\frac{1}{2!}[\hat{\phi},[\hat{\phi},\hat{r}]]
+\cdots ) = \hat{r}^2-\frac{1}{2!}\theta^2 \hat{r}^{-2}
+ \cdots.
\end{eqnarray}
Then one can readily check
how the two commutation relations
\eqref{nccan} and \eqref{ncrdouble} are different:
Using the commutation relation \eqref{nccan} we have
\begin{eqnarray}
\label{lapp}
[\hat{x}^2+\hat{y}^2 ,\hat{x}]
=[\hat{y}^2,\hat{x}]=-2i\theta \hat{y}=-2i\theta \hat{r}\sin\hat{\phi},
\end{eqnarray}
and using \eqref{r2xy} this can be rewritten as
\begin{eqnarray}
\label{rapp}
[\hat{r}^2 + \mathcal{O}(\theta^2),\hat{x}] \cong
[\hat{r}^2,\hat{r}\cos\hat{\phi}]
=\hat{r}[\hat{r}^2,\cos\hat{\phi}] = -2i \theta \hat{r}\sin\hat{\phi},
\end{eqnarray}
where the relation $[\hat{r}^2,\hat{\phi}]=2i\theta$ is applied.
Therefore,
\eqref{nccan} and \eqref{ncrdouble} are equivalent up to first order in $\theta$
and became different from the second order in $\theta$.

Here, we make a short remark about the commutation relation used in \cite{Kim:2007nx}.
There a noncommutative BTZ solution was worked out in the polar coordinates
with the following commutation relation:
\begin{equation}
\label{ncr1}
[\hat{r},\hat{\phi}]= i \theta .
\end{equation}
If we assume that the usual relationship \eqref{rec-pol}
between the rectangular and polar coordinate systems
still  holds in the noncommutative case, then
we get the following relation by applying the commutation relation \eqref{ncr1}:
\begin{eqnarray}
[\hat{x},\hat{y}] = [ \hat{r} \cos{\hat{\phi}},\hat{r} \sin{\hat{\phi}} ]
             = i\theta \hat{r},
\end{eqnarray}
which shows that the commutation relations \eqref{nccan} and  \eqref{ncr1}
are not equivalent even by the dimensional count.


\subsection{Twist perspective}

Here, we prefer to use the commutation relation $[\hat{r}^2,\hat{\phi}]=2i\theta$
in solving the Seiberg-Witten equation for calculational convenience,
since the two commutation relations \eqref{ncr2} and \eqref{ncrdouble} are
exactly equivalent.
The reason for this preference can be easily understood
if we view the Moyal product from the twist perspective.

It is known that the Moyal product \eqref{moyalprd} can also be reproduced
from the deformed $*$-product \cite{Chaichian:2004za,Bu:2006ha}:
\begin{equation}
\label{fstarg}
(f * g)(x)\equiv
\cdot \left[ \mathcal{F}^{-1}_{*}(f(x)\otimes g(x))\right] ,
\end{equation}
where
the multiplication $\cdot$ is defined as $\cdot[f(x)\otimes g(x)] = f(x)g(x)$,
and the twist element $\mathcal{F}_*$ is represented with the
generators of translation along the $x^\alpha$ directions,  $P_\alpha$, as follows.
\begin{equation}
 \mathcal{F_*} =
 e^{\frac{i}{2}\theta^{\alpha\beta}P_\alpha\otimes P_\beta}
~~ \rightarrow~~
e^{-\frac{i}{2}\theta^{\alpha\beta} \frac{\partial}{\partial x^\alpha}
 \otimes\frac{\partial}{\partial x^\beta}}.
 \label{Telement}
\end{equation}
Using \eqref{fstarg} and \eqref{Telement},
one can check that $f*g$ in \eqref{fstarg} is indeed equivalent to
the Moyal product $f\star g$ given in \eqref{moyalprd}:
\begin{eqnarray}
\label{twistpd}
(f* g)(x)&=&\cdot \left[
e^{\frac{i}{2}\theta^{\alpha\beta} \frac{\partial}{\partial x^\alpha}
 \otimes\frac{\partial}{\partial x^\beta}}
 (f(x)\otimes g(x))\right] \nonumber\\
 &=&\cdot \left[ f(x)\otimes g(x)
 +\frac{i}{2}\theta^{\alpha\beta} \frac{\partial f(x)}{\partial x^\alpha}
 \otimes \frac{\partial g(x)}{\partial x^\beta}+\cdots
 \right] \nonumber\\
&=&f(x)\cdot g(x)+\frac{i}{2}\theta^{\alpha\beta} \frac{\partial f(x)}{\partial x^\alpha}
 \frac{\partial g(x)}{\partial x^\beta}+\cdots
 \nonumber\\
 &=&\left.\exp\left[\frac{i}{2}\theta^{\alpha\beta}\frac{\partial}
{\partial x^{\alpha} }\frac{\partial}{\partial y^{\beta} }\right] f(x)g(y)\right|_{x=y}
\nonumber\\
&\equiv&(f\star g)(x) .
\end{eqnarray}
Thus knowing the twist element in a given coordinate system helps one to
 identify the corresponding Moyal product.

The twist element which yields
the noncommutativity \eqref{nccan} in the rectangular coordinates, or $[x,y]_*=i\theta$,
is given by
\begin{equation}
 \mathcal{F_*} =
 \exp\left[-\frac{i\theta}{2} \left(\frac{\partial}{\partial x}\otimes\frac{\partial}{\partial y}
 -\frac{\partial}{\partial y}\otimes\frac{\partial}{\partial x}\right)\right].
 \label{xyelement}
\end{equation}
One can rewrite the above exponent
up to first order in $\theta$,
as follows:
\begin{eqnarray}\label{axialexponent}
\frac{\partial}{\partial x}\otimes\frac{\partial}{\partial y}
 -\frac{\partial}{\partial y}\otimes\frac{\partial}{\partial x}
 & \simeq & \frac{\partial}{\partial r}\otimes \frac{1}{r}\frac{\partial}{\partial \phi}
 -\frac{1}{r}\frac{\partial}{\partial \phi}\otimes\frac{\partial}{\partial r}.
\end{eqnarray}
We can also define the twist element  $\mathcal{F'_*}$
in the polar coordinates
which yields the commutation relation
 $[r,\phi]_{*}=i \theta / r$ as in \eqref{ncr2}
 and is equivalent to $\mathcal{F_*}$ up to first order in $\theta$:
\begin{eqnarray}
 \mathcal{F'_*} & :=  &
 \exp\left[-\frac{i\theta}{2}\left( \frac{1}{r}\frac{\partial}{\partial r}\otimes
 \frac{\partial}{\partial \phi}
 -\frac{\partial}{\partial \phi}\otimes \frac{1}{r}\frac{\partial}{\partial r}\right)\right]
  \label{relement} \\
 & \simeq &
 \exp\left[-\frac{i\theta}{2}\left(\frac{\partial}{\partial r}\otimes\frac{1}{r}
 \frac{\partial}{\partial \phi}
 -\frac{1}{r}\frac{\partial}{\partial \phi}\otimes\frac{\partial}{\partial r}\right)\right]
 \simeq \mathcal{F_*}.
\nonumber
\end{eqnarray}

If we write a twisted product corresponding to $\mathcal{F'_*}$,
it would look like the Moyal product \eqref{moyalprd} except that $\theta$ becomes coordinate dependant,
i.e., $\theta \rightarrow \theta/r$.
To use the solution of the Seiberg-Witten equation
without any modification,
one should carefully place the factor $1/r$
in front of the derivative $\frac{\partial}{\partial r}$~
 in  \eqref{relement}
when one expands the Moyal products
in the Seiberg-Witten equation.
However, if we rewrite $\mathcal{F'_*}$  in terms of the derivative
$\frac{\partial}{\partial r^2}$,
as a new twist element $\mathcal{F''_*}$,
\begin{equation}
 \mathcal{F''_*} =
 \exp\left[-i\theta \left(\frac{\partial}{\partial r^2}\otimes\frac{\partial}{\partial \phi}
 -\frac{\partial}{\partial \phi}\otimes\frac{\partial}{\partial r^2}\right)\right],
 \label{r2element}
\end{equation}
this would allow us to use the Seiberg-Witten relation without any modification.
The new twist element $\mathcal{F''_*}$ is equivalent to  $\mathcal{F'_*}$
and yields the commutation relation
$[r^2,\phi]_{*} =  r^2*\phi-\phi*r^2 = 2i\theta$.



\section{BTZ black hole}
\label{secBTZ}

Here and in the following section
we investigate the effect of non-exact equivalence in noncommutativity
using the two known commutative solutions in 3D,
the BTZ black hole solution \cite{Banados:1992wn,Carlip:1994hq}
and the conical solution on  $AdS_3$  \cite{Pinzul:2005ta},
in two ways.

One way is like the following:
To apply the Seiberg-Witten map associated with
the noncommutativity in  the rectangular coordinates
we first transform the commutative  solution obtained in the polar coordinates
into the one in the rectangular coordinates.
Then after getting noncommutative solutions by applying the Seiberg-Witten map
with the canonical commutation relation of the rectangular coordinates,
we rewrite them back into the polar coordinates.
The other way is to use the Seiberg-Witten map directly
in the polar coordinates without rewriting the solution back and forth
between the polar and the rectangular coordinate systems.
%

The action of the $(2+1)$ dimensional noncommutative $U(1,1)\times U(1,1)$
 Chern-Simons theory with the negative
cosmological constant $\Lambda=-1/l^2$ is given by up to boundary terms \cite{Banados:2001xw,Cacciatori:2002gq},
\begin{eqnarray}
\label{action}
&&\hat{S}(\mathcal{\hat{A}}^{+},\mathcal{\hat{A}}^{-})=
\hat{S}_{+}(\mathcal{\hat{A}}^{+})-\hat{S}_{-}(\mathcal{\hat{A}}^{-}), \\
&& \hat{S}_{\pm}(\mathcal{\hat{A}}^{\pm})=
\beta\int \rm Tr(\mathcal{\hat{A}}^{\pm} \sw d\mathcal{\hat{A}}^{\pm}+\frac{2}{3}
\mathcal{\hat{A}}^{\pm}\sw \mathcal{\hat{A}}^{\pm} \sw \mathcal{\hat{A}}^{\pm}),\nonumber
\end{eqnarray}
where $\beta=l/16\pi G_{N}$ and  $G_{N}$ is the three dimensional Newton constant.
Here
$
\mathcal{\hat{A}^{\pm}}=\mathcal{\hat{A}}^{A\pm}\tau_{A}
=\hat{A}^{a\pm}\tau_{a}+\hat{B}^{\pm}\tau_{3},
$
with  $A=0,1,2,3$,  $~ a={0,1,2},$ ~ $\mathcal{\hat{A}}^{a\pm}=\hat{A}^{a\pm}$,
 $~ \mathcal{\hat{A}}^{3\pm}=\hat{B}^{\pm}$,
 and the deformed wedge product $\sw$ denotes  that
$
A \sw B \equiv A_{\mu} \star B_{\nu}~dx^{\mu} \wedge dx^{\nu}.
$
The noncommutative $SU(1,1) \times SU(1,1)$ gauge fields $\hat{A}$ are expressed in terms of the triad
$\hat{e}$ and the spin connection $\hat{\omega}$ as %
$\label{nc_cs3grav}
 \hat{A}^{a\pm}:=\hat{\omega}^{a}\pm \hat{e}^{a}/{l}.
$
In terms of $\hat{e}$ and  $\hat{\omega}$
the action becomes \cite{Cacciatori:2002gq}
\begin{eqnarray}
\label{reaction}
\hat{S}\!\!&=&\!\!\frac{1}{8\pi G_{N}}\int\left(\hat{e}^{a}\sw \hat{R}_{a}
+\frac{1}{6l^2}\epsilon_{abc}\hat{e}^{a}\sw\hat{e}^{b}
\sw\hat{e}^{c}\right)
\nonumber \\
\!\!&-&\!\! \frac{\beta}{2}
\int\left(\hat{B}^{+}\sw d\hat{B}^{+}
+\frac{i}{3}\hat{B}^{+}\sw\hat{B}^{+}
\sw\hat{B}^{+}\right)
+\frac{\beta}{2}
\int\left(\hat{B}^{-}\sw d\hat{B}^{-}
+\frac{i}{3}\hat{B}^{-}\sw \hat{B}^{-}
\sw\hat{B}^{-}\right)
\nonumber \\
&+&\frac{i\beta}{2} \int( \hat{B}^{+}-\hat{B}^{-})\sw
\left(\hat{\omega}^{a}\sw \hat{\omega}_{a}+\frac{1}{l^2}\hat{e}^{a}
\sw\hat{e}_{a}\right)
\nonumber \\
&+&\frac{i\beta}{2l}\int( \hat{B}^{+}+\hat{B}^{-})\sw
\left(\hat{\omega}^{a}\sw \hat{e}_{a}+\hat{e}^{a}
\sw \hat{\omega}_{a}\right),
\end{eqnarray}
up to surface terms, where  $\hat{R}^{a}=d\hat{\omega}^{a}
+\frac{1}{2}\epsilon^{abc}\hat{\omega}_{b}\stackrel{\star}{\wedge}\hat{\omega}_{c}$.
The equation of motion can be written as follows.
\begin{eqnarray}
\label{nccurtensor}
\hat{\mathcal{F}}^{\pm} \equiv d\hat{\mathcal{A}}^{\pm}+ \hat{\mathcal{A}}^{\pm}
\sw\hat{\mathcal{A}}^{\pm}=0.
\end{eqnarray}
%
In the commutative limit this becomes,
\begin{eqnarray}
\label{ccurtensor}
F^{\pm} \equiv d A^{\pm}+ A^{\pm}\wedge A^{\pm}=0, ~~ d B^{\pm}= 0,
\end{eqnarray}
and the first one can be rewritten as
\begin{equation}
R^{a} + \frac{1}{2l^2}\epsilon^{abc}e_{b}\wedge e_{c}=0, ~~
T^{a} \equiv de^{a}+\epsilon^{abc}\omega_{b}\wedge e_{c}= 0.
\end{equation}
The solution of the decoupled EOM for $SU(1,1)\times SU(1,1)$ part
was obtained in \cite{Carlip:1994hq}:
\begin{eqnarray}
\label{triad}
e^{0}&=& m\left(\frac{r_{+}}{l}dt-r_{-}d\phi\right),~
e^{1}=\frac{l}{n}dm,~
e^{2}=n\left(r_{+}d\phi-\frac{r_{-}}{l}dt\right), \nonumber
\\
\label{spinc}
\omega^{0}&=& -\frac{m}{l}\left(r_{+}d\phi-\frac{r_{-}}{l}\right),~
\omega^{1}=0,~~~~~
\omega^{2}=-\frac{n}{l}\left( \frac{r_{+}}{l}dt-r_{-}d\phi\right),
\end{eqnarray}
where $m^2=(r^2-r_{+}^2)/(r_{+}^2-r_{-}^2)$,~ $n^2=(r^2-r_{-}^2)/(r_{+}^2-r_{-}^2)$,
and $r_+,~ r_-$ are the outer and inner horizons respectively.
There it was also shown to be equivalent to the ordinary BTZ black hole solution \cite{Banados:1992wn}:
\begin{equation}
ds^2=-N^2dt^2+N^{-2}dr^2+r^2(d\phi+N^{\phi}dt)^2,
\end{equation}
where $N^2=(r^2-r_{+}^2)(r^2-r_{-}^2)/l^2 r^2$ and $N^{\phi}=-r_{+}r_{-}/lr^2$.

\subsection{Rectangular coordinates}

The BTZ solution in the polar coordinates can be rewritten
 in the rectangular coordinates as follows:
\begin{eqnarray}
ds^2 &=& [-N^2+r^2(N^{\phi})^2]dt^2-2yN^{\phi}dt dx +2x N^{\phi} dt dy
\nonumber \\
&&+\frac{2xy}{r^2}(N^{-2}-1)dxdy
+\frac{1}{r^2}(N^{-2}x^2+y^2)dx^2
+\frac{1}{r^2}(N^{-2}y^2+x^2)dy^2,
\end{eqnarray}
where  $r^2=x^2+y^2$, ~$r_+^2=\frac{Ml^2}{2}\left\{
1+\left[1-\left(\frac{J}{Ml}\right)^2\right]^{1/2}
\right\}$, ~ $r_-=Jl/2r_+$
, ~$N^{\phi}=-r_{+}r_{-}/lr^2$,
and $N^2=(r^2-r_{+}^2)(r^2-r_{-}^2)/l^2 r^2$.
As in \cite{EDY:20081},
 we consider two simple $U(1)$ fluxes $B_{\mu}^{\pm}=B d\phi=B(xdy-ydx)/r^2$ with constant $B$.
Then, the commutative $U(1,1) \times U(1,1)$ gauge fields $\mathcal{A}^{\pm}$ can be written as
\begin{eqnarray}
\label{cugrectang}
\mathcal{A}^{\pm}_{\mu}= \mathcal{A}^{\pm A}\tau_{A}=
A_{\mu}^{a\pm}\tau_{a}
+B_{\mu}^{\pm}\tau_{3},
\end{eqnarray}
where $A={0,1,2,3}, a={0,1,2},~\mathcal{A}_{\mu}^{a\pm}=A_{\mu}^{a\pm}$,
$\mathcal{A}_{\mu}^{3\pm}=B_{\mu}^{\pm}$ and the gauge fields $A^{a\pm}$ are given by
\begin{eqnarray}
\label{csurectang}
A^{0\pm} &=& \pm \frac{m(r_{+}\pm r_{-})}{l^2}\left[dt \pm \frac{l}{r^2}(ydx-xdy)\right],
\nonumber \\
A^{1\pm}&=& \pm \frac{1}{\sqrt{(r^2-r_{+}^2)(r^2-r_{-}^2)}}(xdx+ydy),
\nonumber \\
A^{2\pm} &=& -\frac{n(r_{+}\pm r_{-})}{l^2}\left[dt \pm \frac{l}{r^2}(ydx-xdy)\right].
\end{eqnarray}
 From the commutative $U(1,1) \times U(1,1)$ gauge fields, we get $\mathcal{A'}_{\mu}^{\pm}$
(recall that $\hat{\mathcal{A}}^{\pm}=\mathcal{A}_{\mu}^{\pm}+\mathcal{A'}_{\mu}^{\pm}$)
via the Seiberg-Witten map \eqref{Aswef} :
\begin{eqnarray}
\mathcal{A'}_{t}^{\pm}
&=& \frac{i \theta B}{8l^2 \sqrt{(r^2-r_{+}^2)(r^2-r_{-}^2)}}
\sqrt{\frac{r_{+}\pm r_{-}}{r_{+} \mp r_{-}}}
\left(
      \begin{array}{cc}
        \mp \sqrt{r^2-r_{-}^2} & -\sqrt{r^2-r_{+}^2} \\
        \sqrt{r^2-r_{+}^2} & \pm \sqrt{r^2-r_{-}^2} \\
      \end{array}
    \right),
   \nonumber \\
\mathcal{A'}_{x}^{\pm}
&=& \frac{i \theta}{8l^2 r^4 (r^2-r_{+}^2)(r^2-r_{-}^2)}
\left(
      \begin{array}{cc}
        -y(U^{\pm}-V^{\mp}) & \pm Bl(yF^{\pm}-ilr^2 x G) \\
        \mp Bl(yF^{\pm}+ilr^2 x G)&  -y(U^{\pm}+V^{\mp}) \\
      \end{array}
    \right),
    \nonumber \\
\mathcal{A'}_{y}^{\pm}
&=& \frac{i \theta}{8l^2 r^4 (r^2-r_{+}^2)(r^2-r_{-}^2)}
\left(
      \begin{array}{cc}
        x(U^{\pm}-V^{\mp}) & \mp Bl(x F^{\pm}+ilr^2 y G) \\
        \mp Bl(x F^{\pm}-ilr^2 y G)&  x(U^{\pm}+V^{\mp}) \\
      \end{array}
    \right),
 \end{eqnarray}
where
\begin{eqnarray}
U^{\pm}&=&(r^2-r_{+}^2)(r^2-r_{-}^2)[B^2l^2-(r_{+}\pm r_{-})]^2-r^4l^2,
\nonumber \\
V^{\mp}&=& B l(r_{+}\mp r_{-})(r^2-2r_{-}^2)\sqrt{(r^2-r_{+}^2)(r^2-r_{-}^2)},
\nonumber \\
F^{\pm}&=& (r^2-r-{+}^2)(r^2-2r_{-}^2)(r_{+}\pm r_{-})\sqrt{\frac{r^2-r_{-}^2}{r_{+}^2-r_{-}^2}},
\nonumber \\
G &=& r^2\sqrt{\frac{r^2-r_{-}^2}{r^2-r_{+}^2}}
-(r^2-2r_{-}^2)\sqrt{\frac{r^2-r_{+}^2}{r^2-r_{-}^2}}.
\end{eqnarray}
Using the relations between
the gauge fields and the triad and spin connection,
 ~ $\hat e/l=\hat{\mathcal{A}}^{+}+\hat{\mathcal{A}}^{-}$
and $ \hat \omega=\hat{\mathcal{A}}^{+}-\hat{\mathcal{A}}^{-}$, we get the following
 up to first order in $\theta$.
\begin{eqnarray}
\label{nctrirectang}
\hat{e}^{0} &=&\frac{r_{+}[r^2-r_{+}^2-\theta B/4]}{l\sqrt{(r^2-r_{+}^2)(r_{+}^2-r_{-}^2)}}dt
+\frac{r_{-}}{r^2}\sqrt{\frac{r^2-r_{+}^2}{r_{+}^2-r_{-}^2}}
\left[ 1+\frac{\theta B}{4r^2}
\left(\frac{r^2-2r_{+}^2}{r^2-r_{+}^2}\right)\right](y dx-x dy),
\nonumber \\
\hat{e}^{1} &=& -\frac{l(r^2+r_{-}^2)}{(r^2-r_{+}^2)(r^2-r_{-}^2)}
\left[ 1-\frac{\theta B}{4r^2}
\frac{r_{+}^4(r^2-2r_{-}^2)-r_{-}^4(r^2-2r_{+}^2)}{(r_{+}^2-r_{-}^2)
(r^2+r_{-}^2)\sqrt{(r^2-r_{+}^2)(r^2-r_{-}^2)}}\right] (xdx+ydy),
\nonumber \\
\hat{e}^{2} &=& \frac{r_{-}[r^2-r_{-}^2-\theta B/4]}{l\sqrt{(r^2-r_{-}^2)(r_{+}^2-r_{-}^2)}}dt
-\frac{r_{+}}{r^2}\sqrt{\frac{r^2-r_{-}^2}{r_{+}^2-r_{-}^2}}
\left[ 1+\frac{\theta B}{4r^2}
\left(\frac{r^2-2r_{-}^2}{r^2-r_{-}^2}\right)\right](y dx-x dy),
\\
\label{ncspincrectang}
\hat{\omega}^{0} &=&\frac{r_{-}[r^2-r_{+}^2-\theta B/4]}{l^2\sqrt{(r^2-r_{+}^2)(r_{+}^2-r_{-}^2)}}dt
+\frac{r_{+}}{lr^2}\sqrt{\frac{r^2-r_{+}^2}{r_{+}^2-r_{-}^2}}
\left[ 1+\frac{\theta B}{4r^2}
\left(\frac{r^2-2r_{+}^2}{r^2-r_{+}^2}\right)\right](y dx-x dy),
\nonumber \\
\hat{\omega}^{1} &=& 0,
\nonumber \\
\hat{\omega}^{2} &=& -\frac{r_{+}[r^2-r_{-}^2-\theta B/4]}{l^2\sqrt{(r^2-r_{+}^2)(r_{+}^2-r_{-}^2)}}dt
-\frac{r_{-}}{lr^2}\sqrt{\frac{r^2-r_{-}^2}{r_{+}^2-r_{-}^2}}
\left[ 1+\frac{\theta B}{4r^2}
\left(\frac{r^2-2r_{-}^2}{r^2-r_{-}^2}\right)\right](y dx-x dy). \nonumber
\end{eqnarray}

A noncommutative length element can be defined by
\begin{eqnarray}
\label{ncmetricrectang}
d\hat{s}^2=\hat{g}_{\mu\nu}dx^{\mu}dx^{\nu} \equiv \eta_{ab}\hat{e}_{\mu}^{a}\star \hat{e}_{\nu}^{b}dx^{\mu}dx^{\nu},
\end{eqnarray}
where $\star$ denotes the Moyal product.
Since  the length element $d\hat{s}^2$ in (\ref{ncmetricrectang}) has symmetric summation,
we end up with a real length element.
Thus we define a real noncommutative metric by
 $\hat{G}_{\mu\nu} \equiv (\hat{g}_{\mu\nu}+\hat{g}_{\nu\mu})/2$
 as in \cite{Pinzul:2005ta}.
After transforming it back to the polar coordinates,  the length element is given by
\begin{eqnarray}
d\hat{s}^2 &=& \hat{G}_{\mu\nu}dx^{\mu}dx^{\nu}
\nonumber \\
&=& -\mathcal{F}^2 dt^2+\mathcal{\hat{N}}^{-2}dr^2
+2r^2 N^{\phi}\left(1+\frac{\theta B}{2r^2}\right)dt d\phi
+r^2\left(1+\frac{\theta B }{2r^2}\right)d\phi^2,
\end{eqnarray}
where
\begin{eqnarray}
\mathcal{F}^2&=&\frac{(r^2-r_{+}^2-r_{-}^2)}{l^2}-\frac{\theta B}{2l^2}=f^2,
\\
\hat{\mathcal{N}}^2&=&\frac{1}{l^2 r^2}\left[ (r^2-r_{+}^2)(r^2-r_{-}^2)
-\frac{\theta B}{2r^2}\left( r_{+}^2(r^2-r_{-}^2)+r_{-}^2(r^2-r_{+}^2)
\right)\right].
\end{eqnarray}

Now, we investigate the apparent and Killing horizons of the above solution by the following relations:
\begin{eqnarray}
\label{apparenth}
\hat{G}^{rr}=\hat{G}_{rr}^{-1}=\hat{\mathcal{N}}^2=0,
\end{eqnarray}
for the apparent horizon (denoted as $\hat{r}$), and
\begin{equation}
 \hat{\chi}^2=
\hat{G}_{tt}-\hat{G}_{t\phi}^2/\hat{G}_{\phi\phi}=0,
\end{equation}
for the Killing horizon (denoted as $\tilde{r}$).
%
These two equations yield the apparent and Killing horizons
up to first order in $\theta$ at
\begin{eqnarray}
\label{apparenth}
\hat{r}_{\pm}^{2}&=&r_{\pm}^{2}+\frac{\theta B}{2}+\mathcal{O}(\theta^2),\\
\label{killingh}
\tilde{r}_{\pm}^2&=&r_{\pm}^2 + \frac{\theta B}{2}
+\mathcal{O}(\theta^2).
\end{eqnarray}

Here the apparent and Killing horizons coincide, and
the inner and outer horizons are shifted from the classical(commutative case) value
by the same amount $\theta B/2$ due to noncommutative effect of flux.
Note that this feature agrees with the result in the commutative(classical) case,
in which the apparent and Killing horizons coincide for stationary black holes.

\subsection{Polar coordinates }
Here, we recall the solution in the noncommutative polar coordinates
obtained in \cite{EDY:20081} for comparison.
 From the consideration in section \ref{secDiff},
the Moyal ($\star$) product from $[\hat{R}, \hat{\phi}]= 2 i \theta$
is given by
\begin{eqnarray}
\label{starp}
(f\star g)(x)=\left.\exp\left[i\theta\left(\frac{\partial}{\partial R}\frac{\partial}{\partial \phi'}
-\frac{\partial}{\partial \phi}\frac{\partial}{\partial R'}\right)\right]f(x)g(x')
\right|_{x=x'},
\end{eqnarray}
 where $\hat{R}\equiv \hat{r}^2$.
 The noncommutative solution $\mathcal{\hat{A}}^{\pm}$ is given by
\begin{eqnarray}
\label{ncgauges}
\mathcal{\hat{A}}^{\pm}_{\mu}=\hat{A}_{\mu}^{a\pm}\tau_{a}+\hat{B}_{\mu}^{\pm}\tau_{3}
= \left(A_{\mu}^{a\pm}-\frac{\theta}{2}B_{\phi}^{\pm}\partial_{R}A_{\mu}^{a\pm}\right)\tau_{a}
+B_{\mu}^{\pm}\tau_{3}+\mathcal{O}(\theta^2),
\end{eqnarray}
where we also considered two $U(1)$ fluxes $B_{\mu}^{\pm}=B d\phi ~$ with constant $B$.

Then from the Sieberg-Witten map
we obtain the noncommutative triad and spin connection as follows.
\begin{eqnarray}
\label{nctriad}
\hat{e}^{0}&=& \left(m-\frac{\theta B}{2}m'\right)\left(\frac{r_{+}}{l}dt-r_{-}d\phi\right)+\mathcal{O}(\theta^2),
\nonumber \\
\hat{e}^{1}&=& l \left[\frac{m'}{n}-\frac{\theta B}{2}\left(\frac{m'}{n}\right)'\right]dR+\mathcal{O}(\theta^2),
\nonumber \\
\hat{e}^{2}&=& \left(n-\frac{\theta B}{2}n'\right)\left(r_{+}d\phi-\frac{r_{-}}{l}dt\right)+\mathcal{O}(\theta^2),
\\
\label{ncspinc}
\hat{\omega}^{0}&=& -\frac{1}{l}\left(m-\frac{\theta B}{2}m'\right) \left(r_{+}d\phi-\frac{r_{-}}{l}\right)+\mathcal{O}(\theta^2),
\nonumber \\
\hat{\omega}^{1}&=&\mathcal{O}(\theta^2),
\nonumber \\
\hat{\omega}^{2}&=&-\frac{1}{l} \left(n-\frac{\theta B}{2}n'\right)
\left( \frac{r_{+}}{l}dt-r_{-}d\phi\right)+\mathcal{O}(\theta^2), \nonumber
\end{eqnarray}
where ${}'$ denotes the differentiation with respect to $R=r^2$.
It should be noted that in the polar coordinates we get a real metric,
$\hat{e}_{\mu} \star \hat{e}_{\nu}=\hat{e}_{\mu}\hat{e}_{\nu}$.
Rewriting $R$ back to $r^2$, we get
\begin{eqnarray}
\label{ncmetric}
d\hat{s}^2=-f^2dt^2+\hat{N}^{-2}dr^2+2r^2 N^{\phi}dtd\phi
+\left(r^2+\frac{\theta B}{2}\right)d\phi^2+\mathcal{O}(\theta^2),
\end{eqnarray}
where
\begin{eqnarray}
N^{\phi}&=&-r_{+}r_{-}/lr^2, \\
f^2&=&\frac{(r^2-r_{+}^2-r_{-}^2)}{l^2}-\frac{\theta B}{2l^2}, \\
\hat{N}^2&=&\frac{1}{l^2 r^2}\left[ (r^2-r_{+}^2)(r^2-r_{-}^2)
-\frac{\theta B}{2}\left(2r^2-r_{+}^2-r_{-}^2\right)\right].
\end{eqnarray}
In this solution, the apparent and Killing horizons denoted as $
\hat{r}$ and $\tilde{r}$, respectively, are given by:
\begin{eqnarray}
\label{apparenth}
\hat{r}_{\pm}^{2}&=&r_{\pm}^{2}+\frac{\theta B}{2}+\mathcal{O}(\theta^2),\\
\label{killingh}
\tilde{r}_{\pm}^2&=&r_{\pm}^2 \pm \frac{\theta B}{2}
\left(\frac{r_{+}^2+r_{-}^2}{r_{+}^2-r_{-}^2}\right)+\mathcal{O}(\theta^2).
\end{eqnarray}
Unlike the rectangular case,
the apparent and the Killing horizons in this case do not coincide.
Note that the outer horizons coincide only in the non-rotating limit in which
 the inner horizon of the commutative solution vanishes($r_{-}=0$).

\section{Conical solution on $AdS_3$}
\label{secCON}

In this section
we first reobtain the noncommutative conical solution
in the rectangular coordinates and check it with the
previously obtained one in \cite{Pinzul:2005ta}.
Then, we repeat the analysis in the polar coordinates
and compare the two results.

\subsection{Rectangular coordinates}

We begin with a nonsingular conical metric on $AdS_3$ in the polar coordinates
$(t,r,\phi)$ \cite{Pinzul:2005ta},
\begin{eqnarray}
\label{ccads}
ds^2=H^{-2}\left[-(2-H)^2(dt+Jd\phi)^2+(1-M)^2 r^2d\phi^2+dr^2\right],
\end{eqnarray}
where $M$, $J$ are mass and angular momentum of the source
 respectively, and $H=(1-r^2/4l^2)$.
The above metric can be transformed to the rectangular coordinates and
the corresponding triad and spin connection in the rectangular coordinates are
given by
\begin{eqnarray}
\label{cadstriad}
e^{0} &=& \frac{2-H}{H}[ dt-\frac{J}{r^2}(y dx-xdy)], \nonumber \\
e^{1} &=& \frac{1}{H}\left[ \left(1-\frac{M y^2}{r^2}\right)dx+\frac{Mxy}{r^2}dy\right], \nonumber \\
e^{2} &=& \frac{1}{H}\left[ \frac{Mxy}{r^2}dx+\left(1-\frac{Mx^2}{r^2}\right)dy\right], \\
\label{cadsspin}
\omega^{0} &=& \frac{(2-M)H-2(1-M)}{H r^2}(xdy-ydx), \nonumber \\
\omega^{1} &=& \frac{y}{l^2 H}\left[ dt-\frac{J}{r^2}(ydx-xdy)\right], \nonumber \\
\omega^{2} &=& -\frac{x}{l^2 H}\left[dt-\frac{J}{r^2}(ydx-xdy)\right]. \nonumber
\end{eqnarray}

As in the previous subsection we consider the same commutative $U(1,1) \times U(1,1)$ gauge fields.
After applying the Seiberg-Witten map we get
$\mathcal{A'}_{\mu}^{\pm}$ as follows.
\begin{eqnarray}
\mathcal{A'}_{t}^{\pm}
&=& \frac{i \theta}{8l^3 H^2}
\left(
      \begin{array}{cc}
        \mp(B+2) & -Bl(2-H)e^{-i\phi}/r \\
        Bl(2-H)e^{-i\phi}/r  & \pm(B-2) \\
      \end{array}
    \right),
   \nonumber \\
\mathcal{A'}_{x}^{\pm}
&=& \frac{i\theta}{8l^2r^3H^2}
\left(
      \begin{array}{cc}
     -ryu^{\pm}_{B}  & \pm iB v^{\pm} \\
      \mp iB \bar{v}^{\pm}  &   -ryu^{\pm}_{-B}\\
      \end{array}
    \right),
    \nonumber \\
\mathcal{A'}_{y}^{\pm}
&=& \frac{i\theta}{8l^2r^3H^2}
\left(
      \begin{array}{cc}
     rxu^{\pm}_{B}  & \pm iB h^{\pm} \\
      \mp iB \bar{h}^{\pm}  &   -rxu^{\pm}_{-B}\\
      \end{array}
    \right),
 \end{eqnarray}
where
\begin{eqnarray}
u^{\pm}_{B}&=&[(M+B)l\pm J]^2-2(1-H)[
J^2\pm 2(M+B+1)Jl+(B^2+2MB+M(M+2)l^2)] \nonumber \\
&& +(1-H)^2[(M-B-2)l \pm J]^2, \nonumber \\
v^{\pm} &=& lx(2-H)+iy(Ml-l\pm J)(3H-2), \nonumber \\
h^{\pm} &=& ly(2-H)-ix(Ml-l\pm J)(3H-2).
\end{eqnarray}
Using the same relations between the gauge fields and the triad and spin connection given
 in the previous section,
we obtain the noncommutative triad and spin connection up to
first order in $\theta$ as follows.
\begin{eqnarray}
\label{nctrirectang}
\hat{e}^{0} &=& \frac{2-H}{H}\left[\left(1-\frac{\theta B}{4l^2 H(2-H)}\right)dt-\frac{J}{r^2}
\left(1-\frac{\theta B}{2r^2}\right)
(y dx-xdy)\right], \nonumber \\
\hat{e}^{1} &=& \frac{1}{r^2 H} \left[Mxy -\frac{\theta B}{16l^2 H}\left(
3(M-1)y^2+\frac{4(M-2)l^2y^2}{r^2}+x^2+4l^2\right)\right]dx \nonumber \\
&& + \frac{1}{H}\left[
\left(1-\frac{Mx^2}{r^2}\right)+\frac{\theta B xy}{16l^2r^4 H} \left(
(2-3M)r^2+4(M-2)l^2\right)\right]dy,
\nonumber \\
\hat{e}^{2} &=& \frac{1}{H}\left[
\left(1-\frac{Mx^2}{r^2}\right)+\frac{\theta B xy}{16l^2r^4 H} \left(
(2-3M)r^2+4(M-2)l^2\right)\right]dx \nonumber \\
&& +\frac{1}{r^2 H} \left[Mxy +\frac{\theta B}{16l^2 H}\left(
3(M-1)x^2-\frac{4(M-2)l^2x^2}{r^2}-y^2+4l^2\right)\right]dy, \\
\label{ncadsspin}
\hat{\omega}^{0} &=& \frac{1}{H r^2}
\left[(2-M)H-2(1-M)-\frac{\theta B}{2r^2}[2(M-1)-(M+2)H]\right](ydx-xdy), \nonumber \\
\hat{\omega}^{1} &=& \frac{y}{l^2 H}\left[1-\frac{\theta B (2-H)}{4r^2 H}\right]dt
-\frac{Jy}{l^2 r^2 H}\left[1-\frac{\theta B(2-3H)}{4r^2 H}\right](ydx-xdy), \nonumber \\
\hat{\omega}^{2} &=& -\frac{x}{l^2 H}\left[1-\frac{\theta B (2-H)}{4r^2 H}\right]dt
+\frac{Jx}{l^2 r^2 H}\left[1-\frac{\theta B(2-3H)}{4r^2 H}\right](ydx-xdy). \nonumber
\end{eqnarray}

Now, the length element of this solution becomes\footnote{
Our conical solution differs from the result
obtained in \cite{Pinzul:2005ta} in one respect, in the use of
 gauge parameter: We use  $\hat g = \hat g(g, A)_{B\neq 0}$ with nonzero flux
 while in \cite{Pinzul:2005ta}
 they used  $\hat g = \hat g(g, A)_{B=0}$ with zero flux.}
\begin{eqnarray}
d\hat{s}^2
&=&-\left( \frac{2-H}{H}\right)^2 \left[ 1-\frac{\theta B}{2l^2 H(2-H)}
\right] dt^2+ H^{-2} \left[ 1-\frac{\theta B}{2r^2} \left(\frac{2-H}{H}\right)
\right]dr^2 \nonumber \\
&& -2J\left( \frac{2-H}{H}\right)^2 \left[ 1+\frac{\theta B}{2r^2}
\left(1-\frac{r^2}{l^2 H(2-H)}\right)\right]dtd\phi \nonumber \\
&& +\frac{1}{H^2} \Bigg[
[(M-1)^2r^2-J^2(2-H)^2]]
\nonumber \\
&& -\frac{\theta B}{2r^2 H} [
2J^2(H^3-6H^2+10H-4)-(M-1)^2r^2(3H-2)]\Bigg]d\phi^2.
\end{eqnarray}
The above solution is not a black hole solution.
However, in order to compare the effect of noncommutativity in different coordinate
systems,
we again consider the same quantities used to evaluate the two horizons,
apparent and Killing horizons in the BTZ black hole case, now denoted as
 $\hat r_{A}$ and $\tilde r_{K}$.
 From the same determining relations,
 $\hat{G}^{rr}=\hat{G}_{rr}^{-1}=0$ and
 $ \hat{\chi}^2=\hat{G}_{tt}-\hat{G}_{t\phi}^2/\hat{G}_{\phi\phi}=0$
 for $\hat r_{A}$ and $\tilde r_{K}$ respectively,
we get
\begin{eqnarray}
\label{apparenth}
{\hat{r}_{A}}^{2}&=&4l^2,\\
\label{killingh}
 \tilde r_{K}^2&=&0,
\end{eqnarray}
up to first order in $\theta$.
The values obtained above coincide with the values in the commutative case.
We consider that
this matches with the feature appeared in the BTZ solution of the rectangular coordinates
given in section 3.1.
There the apparent and Killing horizons coincide
in the noncommutative case just as in the commutative case.

\subsection{Polar coordinates }

Now we do the same analysis in the polar coordinates using $R \equiv r^2$.
 The length element (\ref{ccads}) can be written in the $(t,R,\phi)$ coordinates as
\begin{eqnarray}
\label{ccadspol}
ds^2=H^{-2}\left[-(2-H)^2(dt+Jd\phi)^2+(1-M)^2 R d\phi^2+\frac{dR^2}{4 R}\right].
\end{eqnarray}
Then the triad and spin connection are given by
\begin{eqnarray}
\label{cadstriad}
e^{0} &=& \frac{2-H}{H}(dt+J d\phi), \nonumber \\
e^{1} &=& \frac{1}{H}\left[ \frac{\cos\phi}{2\sqrt{R}}dR-(1-M)\sqrt{R}\sin\phi d\phi\right]
, \nonumber \\
e^{2} &=& \frac{1}{H}\left[ \frac{\sin\phi}{2\sqrt{R}}dR+(1-M)\sqrt{R}\cos\phi d\phi\right] ,
\\
\label{cadsspin}
\omega^{0} &=& \frac{1}{H}[(2-M)H-2(1-M)]d\phi, \nonumber \\
\omega^{1} &=& \frac{\sqrt{R}\sin\phi}{l^2 H}(dt+J d\phi), \nonumber \\
\omega^{2} &=& -\frac{\sqrt{R}\cos\phi}{l^2 H}(dt+J d\phi). \nonumber
\end{eqnarray}


We consider the same $U(1)$ fluxes $B_{\mu}^{\pm}=B d\phi ~$ with constant $B$.
Then, the noncommutative solution
($\mathcal{\hat{A}}^{\pm}=\mathcal{A}^{\pm}_{\mu}+\mathcal{A'}_{\mu}^{\pm}$)
is given by
\begin{eqnarray}
\mathcal{A'}_{t}^{\pm} &=& \mp\frac{i \theta}{8l^3H^2}
\left(
  \begin{array}{cc}
    2+B & \pm Bl(2-H)e^{-i\phi}/\sqrt{R} \\
   \mp Bl(2-H)e^{i\phi}/\sqrt{R} & 2-B \\
  \end{array}
\right),
\nonumber \\
\mathcal{A'}_{R}^{\pm} &=& \pm\frac{\theta B(3H-2)}{16lH^2 R^{3/2}}
\left(
  \begin{array}{cc}
 0 & e^{-i\phi} \\
   e^{i\phi} & 0 \\
  \end{array}
\right),
\nonumber \\
\mathcal{A'}_{\phi}^{\pm} &=& \frac{i \theta(l-Ml\mp J)}{8l^3H^2}
\left(
  \begin{array}{cc}
    2+B & \pm Bl(2-H)e^{-i\phi}/\sqrt{R} \\
   \mp Bl(2-H)e^{i\phi}/\sqrt{R} & 2-B \\
  \end{array}
\right).
\end{eqnarray}
Then using the same relations between the gauge fields and the triad and spin connection
given in the previous section,
 the noncommutative triad and spin connection are given by
\begin{eqnarray}
\label{nctriad}
\hat{e}^{0}&=& \frac{H(2-H)-\theta B/4l^2}{H^2}(dt+J d\phi),
\nonumber \\
\hat{e}^{1}&=& \frac{\cos\phi}{2\sqrt{R}H}\left[ 1+\frac{\theta B}{4R}
\left(\frac{3H-2}{H}\right)\right]dR-\frac{(1-M)\sqrt{R}\sin\phi}{H}\left[
1-\frac{\theta B}{4R}\left(\frac{2-H}{H}\right)\right]d\phi,
\nonumber \\
\hat{e}^{2}&=& \frac{\sin\phi}{2\sqrt{R}H}\left[ 1+\frac{\theta B}{4R}
\left(\frac{3H-2}{H}\right)\right]dR+\frac{(1-M)\sqrt{R}\cos\phi}{H}\left[
1-\frac{\theta B}{4R}\left(\frac{2-H}{H}\right)\right]d\phi, \nonumber
\\
\label{ncspinc}
\hat{\omega}^{0}&=& \frac{1}{H}\left[
(2-M)H-2(1-M)+\frac{\theta B (1-M)}{4l^2 H}\right]d\phi,
\nonumber \\
\hat{\omega}^{1}&=& \frac{\sqrt{R}\sin\phi}{l^2H}\left[
1-\frac{\theta B}{4R}\left(\frac{2-H}{H}\right)\right](dt+Jd\phi),
\nonumber \\
\hat{\omega}^{2}&=& -\frac{\sqrt{R}\cos\phi}{l^2H}\left[
1-\frac{\theta B}{4R}\left(\frac{2-H}{H}\right)\right](dt+Jd\phi).
\end{eqnarray}

 The noncommutative length element defined in the same way as in the previous section
  is  given by in terms of $r$  as follows.
\begin{eqnarray}
\label{ncadsmetric}
d\hat{s}^2
&=& -\hat{\mathcal{F}}^2 dt^2+ \hat{\mathcal{N}}^{-2}dr^2-2J\hat{\mathcal{F}}^2dtd\phi
\nonumber \\
&&+\frac{(1-M)^2r^2-J^2(2-H)^2}{H^2}
\left[ 1-\frac{\theta B}{2l^2}\left(\frac{2-H}{H}\right)\frac{(1-M)^2l^2-J^2}{(1-M)^2r^2-J^2(2-H)^2}
\right]d\phi^2+\mathcal{O}(\theta^2), \nonumber \\
\end{eqnarray}
where
\begin{eqnarray}
\hat{\mathcal{F}}^2 &=& \left(\frac{2-H}{H}\right)^2 \left[1-\frac{\theta B}{2l^2}\frac{1}{H(2-H)}\right], \nonumber \\
\hat{\mathcal{N}}^2 &=& H^2 \left[1-\frac{\theta B}{2r^2}
\left(\frac{3H-2}{H}\right)\right].
\end{eqnarray}

Here we again consider the same quantities $\hat r_{A}$ and $\tilde r_{K}$
defined in the previous subsection to investigate the effect of
noncommutativity in different coordinate systems.
Now they are given by
\begin{eqnarray}
\hat r_{A}^{2}&=&4l^{2}+\mathcal{O}(\theta^2),
\\
\tilde r_{K}^2&=&\frac{\theta B}{4}+\mathcal{O}(\theta^2).
\end{eqnarray}
Unlike the rectangular case in the previous subsection in which
both $\hat r_{A}$ and $\tilde r_{K}$ coincide with the classical values,
here only $\hat r_{A}$ coincides with the classical value $r_A=2l$.
For $\tilde r_{K}$, which would correspond to the Killing horizon of
a black hole, does not coincide with the classical value $r_K=0$.
However, in the non-rotating limit ($J=0$), the solution for $\tilde{r}_K$
does not exist, and this feature agrees with that of the commutative case
in which the solution for $r_K$ does not exist either in the non-rotating limit.
Thus we see that the same pattern holds in the polar coordinates as in the BTZ case, namely
in the non-rotating limit the same feature appears in both commutative and noncommutative cases.

\section{Disscussion}
\label{disscuss}

In this paper, in order to investigate the non-exact equivalence
between noncommutative coordinate systems
we obtain
a noncommutative BTZ black hole solution
in the canonical rectangular coordinates via Seiberg-Witten map,
and compare it with the previously obtained result
in the noncommutative polar coordinates \cite{EDY:20081}.
We repeat the same analysis for the conical solution
in noncommutative $AdS_3$ using the same action
to see whether there exists any
similarity between the two cases.


What we have learned can be illustrated as follows:
\vspace{0.1cm}
\begin{center}
\mbox{\large \xymatrix{ \mathcal{A}(r,\phi)  \ar[dd]_{[\hat r,\hat \phi]=i\tilde{\theta}~~~}^{I}
\ar[rrrr]^{II}
& & & &\mathcal{B}(x,y) \ar[dd]_{III}^{~~~[\hat x,\hat y]=i\theta}
\\
\\
\hat{\mathcal{A}}(r,\phi)   & & & &\ar[llll]^{IV} \hat{\mathcal{B}}(x,y)&, } }
\end{center}
\vspace{0.1cm}
where $\mathcal{B}(x,y)\equiv \mathcal{A}[r(x,y),\phi(x,y)]$
and the maps  $II$, $IV$ are the coordinate transformations
$(x,y)\leftrightarrow (r,\phi)$ in a commutative space,
and the maps  $I$, $III$ denote corresponding Seiberg-Witten maps.
For a function $\mathcal{A}(r,\phi)$,
for example, the Carlip {\it et. al.}'s BTZ black hole solution in the polar coordinates \cite{Carlip:1994hq},
we have two different routes of getting
Seiberg-Witten solutions $\hat{\mathcal{A}}(\mathcal{A})$,
via $I$ or via $II\rightarrow III\rightarrow IV$.
 From the observation of Eq. \eqref{diffSW} in section 2,
we know that the two solutions via the different routes
would be different,
i.e. $\hat{\mathcal{A}}(r,\phi) \neq \hat{\mathcal{B}}[x(r,\phi),y(r,\phi)]$,
since the transformation
$(x,y)\leftrightarrow (r,\phi)$ is not linear.
The results in sections 3 and 4 just support this observation.


Another lesson we get is from the following observations.
1) In the rectangular coordinates, the feature appeared in
the solution of the commutative case remains intact in
the noncommutative case: In the BTZ case, both apparent and
Killing horizons coincide. In the conical solution, the commutative
and the noncommutative results are the same.
2) In the polar coordinates, the feature appeared in the
commutative case is not maintained in the noncommutative case:
In the BTZ case, apparent and Killing horizons do not coincide.
In the conical solution, the commutative
and the noncommutative results do not agree.
However, in the non-rotating limit the feature appeared in
the commutative case is maintained in the noncommutative case:
In the BTZ case, apparent and Killing horizons do coincide.
In the conical solution case, the commutative
and noncommutative results agree.

Thus we are left with a task of understanding the differed behaviors
in the polar coordinates.
Our understanding is as follows.
In the BTZ case, the Killing vector
which determines the Killing horizon
is dependent on the translation generator along the $\hat{\phi}$ direction,
while the apparent horizon is determined by the null vector given by the translation
generator along the radial $\hat{r}$ direction.
Hence in the rotating case the relation between the two horizons is affected by the noncommutativity
between the two coordinates  $(\hat{r},  \hat{\phi})$, and will differ from the commutative case.
The two horizons will not coincide.
In the non-rotating case, the Killing vector does
not depend on the translation generator along the $\hat{\phi}$ direction
and thus no effect of noncommutativity among $(\hat{r},  \hat{\phi})$ enters, resulting
the same relation  as in the commutative case.
In the conical solution case, since we used the same defining relations for $\hat{r}$ and $\tilde{r}$
as in the BTZ case,
we expect the same.

In the rectangular coordinates, the above noncommutative effect does not enter
since we are applying the above operation (getting a solution for $\hat{r}$ and $\tilde{r}$)
to the result obtained by commutative coordinate transformation after the
Seiberg-Witten map, thus wiping out the noncommutative characteristics.
Note that the result obtained in the rectangular coordinates for the BTZ case differs
from the commutative result. However, the feature that the apparent and Killing horizons
coincide remains the same as in the commutative case.
Namely, we simply obtained a differed geometry from the commutative case
due to noncommutative effect by the Seiberg-Witten map. However, the
noncommutative effect in getting the solution of $\hat{r}$ and $ \tilde{r}$
was lost.

Thus as it was pointed out in \cite{ag06} that
the conventional sense of diffeomorphism is not invariant
in noncommutative theory, we
better use the same coordinate system throughout the process of
solution finding, matching the coordinate system such that
the operational meaning of noncommutativity can be kept.
%
%
For instance, the commutation relation $[\hat x,\hat y]=i\theta$
has translational symmetry,
while the commutation relation $[{\hat r}^2,\hat \phi]=2i\theta$
has rotational symmetry.
So if we use $[{\hat r}^2,\hat \phi]=2i\theta$ instead of
$[\hat x,\hat y]=i\theta$, this means that
we choose the rotational symmetry (translational symmetry along $\phi$ direction)
at the cost of the translational symmetry along the x and y directions.
We consider this as the underlying reason for the
differences in the results obtained in the paper.


\section*{Acknowledgments}
This work was supported by the Korea Science and Engineering Foundation(KOSEF) grant
funded by the Korea government(MEST), R01-2008-000-21026-0(E. C.-Y. and D. L.),
and by the Korea Research Foundation grant funded by
the Korea Government(MEST), KRF-2008-314-C00063(Y. L.).

\end{document}